\documentstyle[amssymb,epsfig,prd,aps]{revtex}
\tightenlines
\newcommand\cmp[3]{{\it Comm.\ Math.\ Phys.\ }{\bf #1}, #3 (#2)}

\newcommand\jgp[3]{{\it J.\ Geom.\ Phys.\ }{\bf #1}, #3 (#2)}

\newcommand\npb[3]{{\it Nucl.\ Phys.\ }{\bf B #1}, #3 (#2)}
\newcommand\plb[3]{{\it Phys.\ Lett.\ }{\bf B #1}, #3 (#2)}
\newcommand\pr[3]{{\it Phys.\ Rev.\ }{\bf D #1}, #3 (#2)}
\renewcommand\prl[3]{{\it Phys.\ Rev.\ Lett.\ }{\bf #1}, #3 (#2)}
\newcommand\prep[3]{{\it Phys.\ Rep.\ }{\bf #1}, #3 (#2)}

\newcommand\ijmpa[3]{{\it Int.\ J.\ Mod.\ Phys.\ }{\bf A #1}, #3 (#2)}
\newcommand\sjnp[3]{{\it Sov.\ Jour.\ Nucl.\ Phys.\ }{\bf #1}, #3 (#2)}
\newcommand\npps[3]{{\it Nucl.\ Phys.\ Proc.\ Suppl.\ }{\bf #1}, #3 (#2)}
\newcommand\ptps[3]{{\it Prog.\ Theor.\ Phys.\ Suppl.\ }{\bf #1}, #3 (#2)}

\def\hepth#1{ {\tt hep-th/#1}}

\def\ie{{\it i.e. }}%--- i.e.
\newcommand{\beq}{\begin{equation}}
\newcommand{\eeq}{\end{equation}}
\newcommand{\beqa}{\begin{eqnarray}}
\newcommand{\eeqa}{\end{eqnarray}}
\def\ben{\begin{enumerate}}
\def\een{\end{enumerate}}
\def\ppder#1#2#3{\frac{\partial^2 #1}{\partial #2\partial #3}}

\def\N{{\cal N}}

\def\Tr{{\rm Tr}}

\begin{document}

\title{\bf Phases of dual superconductivity and confinement in softly broken
${\cal N} = 2$ supersymmetric Yang--Mills theories}

\author{Jos\'e D. Edelstein$^{~a}$, Wifredo Garc\'{\i}a Fuertes$^{~b}$, Javier
Mas$^{~a}$, Juan Mateos Guilarte$^{~c}$ \\ \vskip3mm
$~^a$ {\em Departamento de F\'{\i}sica de Part\'{\i}culas, Universidade de
Santiago de Compostela \\
E-15706 ~Santiago de Compostela, Spain} \\ \vskip1mm
$~^b$ {\em Departamento de F\'{\i}sica, Facultad de Ciencias, Universidad de
Oviedo \\ E-33007 ~Oviedo, Spain} \\ \vskip1mm
$~^c$ {\em Departamento de F\'{\i}sica, Facultad de Ciencias, Universidad de
Salamanca \\ E-37008 ~Salamanca, Spain}}

\maketitle

\begin{abstract}
We study the electric flux tubes that undertake color confinement in ${\cal
N}=2$ supersymmetric Yang--Mills theories softly broken down to ${\cal N}=1$
by perturbing with the first two Casimir operators. The relevant
Abelian Higgs model is  not the standard one due to the presence of an
off-diagonal coupling among different magnetic $U(1)$ factors. We perform a
preliminary study of this model at a qualitative level.  BPS vortices are
explicitely obtained for particular values of the soft breaking parameters.
Generically however, even in the ultrastrong scaling limit, vortices are not
critical but live in a ``hybrid'' type II phase. Also, ratios among string
tensions are seen to follow no simple pattern. We examine the situation at
the half Higgsed vacua  and find evidence for solutions with the behaviour of
superconducting strings. In some cases they are solutions to BPS equations.

{Preprint numbers: US-FT/26-99~~~FFUOV-99/13}
\vskip3mm

\end{abstract}

%%%%%%%%%%%%%%%%%%%%%%%%%%%%%%%%%%%%%%%%%%%%%%%%%%%%%%%%%%%%%%%%%%%%%%%%%%%%%%%
%%%%%%%%%%%                         Section I                     %%%%%%%%%%%%%
%%%%%%%%%%%%%%%%%%%%%%%%%%%%%%%%%%%%%%%%%%%%%%%%%%%%%%%%%%%%%%%%%%%%%%%%%%%%%%%
\section{Introduction}

Certainly, one of the most beautiful ideas in the context of quantum
chromodynamics (QCD) is the confinement mechanism envisaged by 't Hooft
\cite{tHooft} and Mandelstam \cite{Mandel} through the condensation of light
monopoles. In essence it states that the QCD vacuum should behave as a {\em
dual} superconductor where magnetic order takes place, and electric flux tubes
form  thus producing color confinement. In the context of QCD it stands for a
kind of descriptive scheme, as long as it is not known how magnetically charged
quanta can arise and condense in the effective low energy theory. In this
respect, the idea of Abelian projection proposed by 't Hooft has received
increasing support from numerical simulations on the lattice in the last few
years \cite{Lattice}. Even in the continuum, recent work using a novel
parametrization of QCD \cite{FadNie}, points in the direction of the above
scenario for color confinement \cite{Cho}. From the analytical side, the
understanding of non-perturbative phenomena in four dimensional quantum field
theory has been put several steps forward since Seiberg and Witten constructed
an exact solution for the low energy dynamics of $SU(2)$ ${\cal N}=2$
supersymmetric Yang-Mills theory \cite{SeiWitt}. In particular, it was
possible for them to show that the mechanism of color confinement devised by
't Hooft and Mandelstam takes place when supersymmetry is broken down to
${\cal N}=1$. These results were soon extended to the case of $SU(N)$
\cite{Klemmetal,DS,AD}. Furthermore, when ${\cal N}=2$ supersymmetry is softly
broken down to ${\cal N}=0$, the same mechanism has been shown  to persist
\cite{Luisetal}. 

In spite of the fact that these results are well known, not much attention has
been paid to the actual solutions in the strong coupling limit corresponding
to electric flux lines that would undertake quark confinement. In
Ref.\cite{DS}, it was shown that this sort of vortices should have a
spectrum of string tensions that distinguishes among different factors in the
magnetic $U(1)^{N-1}$ theory arising in the infrared. The same result was
found in the framework of the M-theory fivebrane version of QCD,  also named
MQCD \cite{HSZ}. The string tension of the $N-1$ electric flux tubes
$T_k~,~~k=1,...,N-1~,$ is given --up to a dimensionful factor which is
different for each theory but independent of  $k$--, by a dimensionless
function $f_N(k) = \sin(\pi k/N)$. This function is somehow universal
as long as the soft breaking perturbation is carried by a single Casimir
operator \cite{DS,HSZ,Str}.  Even in that case, the problem of finding such
solutions in the particular model that emerges in this context has not yet
been addressed in  detail, \footnote{~Except for  $SU(2)$, both at the maximal
singularity of the Coulomb branch for pure gauge \cite{GFMG}, and on the Higgs
branch in the theory with massive fundamental matter \cite{Yung}.} probably
due to the naive expectation that the effective theory consists of $N-1$
copies of the standard $U(1)$ Abelian Higgs model. It was already shown by
Douglas and Shenker that the magnetic $U(1)$ factors of the infrared quantum
theory describing the neighborhood of the monopole singularities are  coupled
\cite{DS}. The existence of these off-diagonal couplings, $\tau_{ij}^{\rm
off}$, was confirmed in two different frameworks. First, they appear in the
expression of the Donaldson--Witten  functional for gauge group $SU(N)$
\cite{MM}. Moreover, these couplings were shown to satisfy a stringent
constraint coming from the Whitham hierarchy formulation of the
Seiberg--Witten solution in Ref.\cite{edemas} where, in addition, a general
ansatz for $\tau_{ij}^{\rm off}$ is given.

In this paper, we extend the work \cite{GFMG} to the case of $SU(N)$, ${\cal
N}=2$ supersymmetric Yang-Mills theory softly broken to ${\cal N}=1$.  The
analysis is performed in a ``peculiar'' scaling limit (named ``ultrastrong''
in \cite{HSZ}).  We   show that, even in that limit, generically  there are no
BPS electric flux tubes. A perturbative analysis leads to the conclusion that
the  phase of dual superconductivity is of type II \ie  there is a short range
repulsive force between different vortices. This fact supports the  expectation
that indeed electric flux lines are safely confined into stable flux tubes, a
feature of the confinement mechanism which is not {\em a priori} granted.
\footnote{~It could happen, for example, that the electric vortices result to
be unstable, and their core grows and smears in such a way that they do not
lead to confinement of electric charges \cite{AdRH}.} It is worth mentioning
in this respect that numerical simulations in lattice QCD seem to point out
that the type of Abelian Higgs model behind the picture of dual
superconductivity is a critical one between type I and type II \cite{BSS}. 

The plan of the paper is as follows. The setup of the problem is given in
Section II where  some aspects of the low-energy dynamics of ${\cal N}=2$
supersymmetric gauge theories softly broken down to ${\cal N}=1$ are reviewed.
We emphasize the existence of non-vanishing couplings between the different
$U(1)$ factors --even at the maximal singularity--, which play an essential
r\^ole in our results. In Section III, we show that the string tension of
vortex-like configurations obeys a Bogomol'nyi bound in the  ultrastrong
scaling limit. However, there are no BPS electric vortices in the system
unless the complex phases of the soft breaking parameters corresponding to
different Casimir operators are aligned. Even in this case, we show that the
string tensions of the resulting BPS vortices are governed by a dimensionless
function $f_N(k)$ which is different from the one obtained in \cite{DS,HSZ},
the latter being recovered as a particular limit of our system corresponding
to a single quadratic ${\cal N}=1$ perturbation. In Section IV, we focus for
convenience on the group $SU(3)$  and analyze the  critical vortex solutions
in certain simplified cases. We speculate about the full spectrum of such
configurations. A perturbative analysis of the dynamics expected for nearly
critical vortices is performed in sections V and VI by means of energetic
arguments. This analysis reveals the existence of repulsive forces among
vortices corresponding to different magnetic $U(1)$ factors. Thereafter we
refer to this phase as an ``hybrid'' type II phase.  In Section VII, the 
half Higgsed vacua are considered. The similarities and  differences with the
model proposed by Witten to describe cosmic superconducting strings 
\cite{Witten} are discussed.  We find solutions to the Bogomol'nyi equations
with the behavior of superconducting strings. Finally, Section VIII is
devoted to our conclusions and further remarks.

%%%%%%%%%%%%%%%%%%%%%%%%%%%%%%%%%%%%%%%%%%%%%%%%%%%%%%%%%%%%%%%%%%%%%%%%%%%%%%%
%%%%%%%%%%%                         Section II                     %%%%%%%%%%%%
%%%%%%%%%%%%%%%%%%%%%%%%%%%%%%%%%%%%%%%%%%%%%%%%%%%%%%%%%%%%%%%%%%%%%%%%%%%%%%%
\section{Infrared dynamics at maximal singularities}

The quantum moduli space of vacua ${\cal M}_\Lambda$ of $SU(N)$, ${\cal N}=2$
supersymmetric gauge theory has a singular locus given by  hypersurfaces of
complex codimension one which may intersect with each other \cite{Klemmetal}.
Along each of these hypersurfaces, an extra massless degree of freedom --whose
quantum numbers can be read off from the monodromy matrix corresponding to a
closed path encircling the singularity--, must be included into the effective
action. At the intersections, many states become simultaneously massless. Of
special interest are those singularities where $N-1$, \ie the maximum allowed
number of mutually local states, become massless. They are accordingly called
maximal singularities. \footnote{~In $SU(3)$, for example, the singular
locus is given by the complex curves
$$
4u^3 - 27(v \pm 2\Lambda^3)^2 = 0 ~,
$$
where $u = 1/2\langle  \Tr\phi^2 \rangle$ and $v = 1/3\langle \Tr\phi^3 
\rangle$
are the gauge  invariant order parameters constructed out of the scalar field
belonging to the ${\cal N}=2$ vector supermultiplet, and $\Lambda$ is the
quantum dynamical scale. Higher intersections of these curves lead to the
so-called $Z_2$ and $Z_3$ singularities, given respectively by  the  points
$\{u^3 = 27\Lambda^6, v = 0 \}$ and   $\{u = 0, v^2 = 4\Lambda^6\}$.}

The addition of a microscopic superpotential breaks supersymmetry and
leads to  an  ${\cal N}=1$ theory
\beq
W_{{\cal N}=1} = \sum_{k=2}^N \frac{1}{k} ~\lambda_k~\Tr\Phi^k ~.
\label{superpot}
\eeq
 Notice that those contributions in (\ref{superpot}) with $k>3$ are
non-renormalizable. However, this does not necessarily mean that they do not
affect the low-energy dynamics. They could be dangerously irrelevant operators
\cite{KSS,emm}. We will not discuss these subtleties here, and shall
restrict ourselves to the case of up to cubic perturbations, 
\beq
W_{{\cal N}=1} = \frac{1}{2} ~\mu~\Tr\Phi^2 + \frac{1}{3} ~\nu~\Tr\Phi^3 ~.
\label{superpoten}
\eeq 
This breaking is soft, in fact renormalizable. The continuum vacuum
degeneracy is lifted except for a given set of points which depend on the
actual values of the parameters $\mu$ and $\nu$. \footnote{~For instance, in
the case of $SU(3)$, the theory has generically five ${\cal N}=1$ vacua,
three of which are the maximal $Z_2$ points. In the limit $\mu\to 0$, the
remaining vacua approach the $Z_3$ points \cite{AD}.} 

Let us focus on the low-energy effective field theory near a maximal point
that we choose, for simplicity, to be that with real quadratic Casimir, $u =
N\Lambda^2$. This is a dual ${\cal N}=2$ supersymmetric gauge system with
gauge group $U(1)^{N-1}$ which includes both chiral multiplets
$\Psi_i^D = (\chi^D_i,V^D_i)$, as well as hypermultiplets $H_i =
(M_i,\tilde{M}_i)$ that correspond to the monopoles that become light in that
patch of the moduli space. One can choose a homology basis for the cycles on
the auxiliary curve such that each monopole has a unit of charge with respect
to each dual gauge field. The quantities $\chi^D_i$, $M_i$ and $\tilde{M}_i$
are chiral ${\cal N}=1$ superfields, while $V^D_i$ are ${\cal N}=1$ vector
superfields (and $W_\alpha^{Di}$ their corresponding superfield strengths).
For completeness, we give also the ${\cal N}=0$ content of these superfields
\begin{eqnarray*}
\chi^D_i=(\phi^D_i,\psi_i,F_i) & ~~~~~~~~~~~ &
V^D_i=((A^D_\mu)_i,\lambda_i,D_i) \\
M_i=(\phi_{m_i},\psi_{m_i},F_{m_i}) & ~~~~~~~~~~~ &
\tilde{M}_i=(\tilde{\phi}_{m_i},\tilde{\psi}_{m_i},\tilde{F}_{m_i})
\end{eqnarray*}
where the notation for fermionic, bosonic and auxiliary components is the
standard one. Setting $a^D_i \equiv \langle \phi^D_i\rangle $, the
coordinates at the point of maximal singularity that we are focusing on are 
$a^D_i= 0$. The dominant piece of the ${\cal N}=2$ low energy effective 
Lagrangian is given in terms of a holomorphic function $ {\cal F}$, called
the effective prepotential
\begin{eqnarray}
{\cal L}_{eff}^{N=2} & = & \frac{1}{4\pi}{\rm Im}\left[\int
d^4\theta~\frac{\partial{\cal F}(\chi^D)}{\partial\chi^D_i}\chi_i^{D\dagger}
+\frac{1}{2}\int d^2\theta~\frac{\partial^2{\cal
F}(\chi^D)}{\partial\chi^D_i\partial\chi^D_j} W_\alpha^{Di} W^{D\alpha
j}\right] \nonumber \\
& + & \int d^4\theta\{M_i^\dagger e^{2V^D_i}M_i + \tilde{M}_i^\dagger
e^{-2V^D_i}\tilde{M}_i\} + {\rm Re}~\int d^2\theta ~W(\chi^D,M,\tilde{M}) ~.
\label{lagn=2}
\end{eqnarray}
The monopole fields  have been ``integrated in'' in order to soak up the
singularity of the effective action when $a^D_i = \langle \phi^D_i\rangle
\to 0$ where $M_i$ becomes massless. The effective superpotential at low
energies is
\beq
W(\chi^D,M,\tilde{M}) =\sqrt{2} M_i\chi^D_i\tilde{M}_i + \mu~{\cal U}(\chi^D)
+ \nu~{\cal V}(\chi^D) ~,
\label{superp}
\eeq
the last two terms being the effective contribution of the supersymmetry
breaking superpotential (\ref{superpoten}). In fact, ${\cal U}$ and
${\cal V}$ are the Abelian superfields arising respectively from the quadratic
and cubic Casimir operators in the low-energy theory. The vacuum expectation
value of their lowest components, $U$ and $V$, are the holomorphic coordinates
in ${\cal M}_\Lambda$, $\langle U\rangle = u$ and $\langle V\rangle = v$.

Written in component fields, the bosonic sector of the system is described by
the Lagrangian
\begin{eqnarray}
{\cal L}_{eff,~B}^{{\cal N}=1} & = & - \frac{1}{4} b_{ij} (F_{\mu\nu})_i
(F^{\mu\nu})_j + (D_\mu\phi_{m_i})^* D^\mu\phi_{m_i} + D_\mu\tilde{\phi}_{m_i}
(D^\mu\tilde{\phi}_{m_i})^* + b_{ij} \partial_\mu{\phi^D_i}^*
\partial^\mu\phi^D_j \nonumber \\
& - & \left[ \frac{1}{2} b_{ij} D_i D_j + b_{ij} F_i^* F_j + F_{m_i}^* F_{m_i}
+ \tilde{F}_{m_i}^* \tilde{F}_{m_i} \right] ~,
\label{efflag}
\end{eqnarray}
where the auxiliary fields are solved as
\begin{eqnarray}
D_i = - b_{ij}^{-1} (|\phi_{m_j}|^2-|\tilde{\phi}_{m_j}|^2) & ~~~~~~~~~~~ &
F_i = - b_{ij}^{-1} \left(\sqrt{2} \phi_{m_j} \tilde{\phi}_{m_j} + C_j\right)
\label{DyF} \\
F_{m_i} = - \sqrt{2}{\phi^D_i}^*\tilde{\phi}_{m_i}^* ~~~~~~~~~ & ~~~~~~~~~~~ &
\tilde{F}_{m_i} = - \sqrt{2}{\phi^D_i}^*{\phi}_{m_i}^* ~,
\label{Efemes}
\end{eqnarray}
whereas field strengths and covariant derivatives are given by
\beqa
(F_{\mu\nu})_i &=& \partial_\mu (A^D_\nu)_i - \partial_\nu (A^D_\mu)_i ~,
\label{fields}
\\
D_\mu\phi_{m_i} = \partial_\mu\phi_{m_i} + i(A^D_\mu)_i\phi_{m_i}
  &\rule{0mm}{6mm}&~~~~~  D_\mu\tilde{\phi}_{m_i} =
\partial_\mu\tilde{\phi}_{m_i} - i(A^D_\mu)_i\tilde{\phi}_{m_i} ~.
\label{covder}
\eeqa
Concerning $C_j$ in (\ref{DyF}), it stands for
\beq
C_j(\phi^D) = \mu U_j(\phi^D) + \nu V_j(\phi^D) \equiv |C_j| e^{i\beta_j} ~,
\label{condens}
\eeq
where $U_j$ and $V_j$ are the derivatives of $U$ and $V$ with respect to
$\phi^D_j$ \cite{DS}, 
\begin{eqnarray}
U_j(\phi^D) = u_j^{(0)} \Lambda + \sum_{p\geq 1} u_j^{(p)}(\phi^D)
\Lambda^{1-p} ~, & ~~~~~~~~ & u_j^{(0)} = -2j \sin\hat\theta_j \label{ui}
\\ V_j(\phi^D) = v_j^{(0)} \Lambda^2 + \sum_{p\geq 1} v_j^{(p)}(\phi^D)
\Lambda^{2-p} ~, & ~~~~~~~~ & v_j^{(0)} = -2j \sin 2\hat\theta_j ~, \label{vi}
\end{eqnarray}
while $u_j^{(p)}(\phi^D)$ and $v_j^{(p)}(\phi^D)$ are homogeneous
polynomials in $\phi^D_i$ of degree $p$, so that $C_j$ are regular functions
in the vicinity of the maximal singularity. Finally, $b_{ij}$ is
($\frac{1}{4\pi}$ times) the imaginary part of the period matrix
$\tau^D_{ij}$,
\beq
\tau^D_{ij}(\phi^D) = \ppder{\cal F}{\phi^D_i}{\phi^D_j} = {1 \over 2\pi i }
\log
\left({\phi^D_i\over \Lambda_i}\right) \delta_{ij} + \tau_{ij}^{\rm off} +
{\cal O}\left(\frac{\phi^D}{\Lambda}\right) ~,
\label{bij}
\eeq
where $\Lambda_j = \Lambda\sin \hat\theta_j $ and $\hat\theta_j = j\pi/N$. 
When expanding around the vacuum expectation value $a^D_i =
\langle\phi^D_i\rangle$, $\tau^D_{ij}(\phi^D)$ yields the effective coupling
constant matrix. The logarithmic singularity when $a^D_i= 0$ corresponds to
the perturbative running of the dual coupling constant up to the maximal
point, displaying the asymptotic freedom of the dual description. The
coupling flows to zero due to the fact that the quantum fluctuations of
massless monopoles have been integrated out. This is fine as long as one is
interested only in searching for  vacuum solutions. Then $M$ and $M^\dagger$
in (\ref{lagn=2})--(\ref{superp}) stand for the zero modes of the monopole
field (see the discussion in \cite{Luisetal}).  Here, however, in order not to
run into double counting of degrees of freedom we should introduce, on
physical grounds, an infrared cut off for the monopole loop integrals. In each
$U(1)$ factor the natural energy scale is set by  the soft breaking
parameters $a^D_i \sim  | C_i^{(0)}| ^{1/2}$ with 
\beq
C_i^{(0)} = \mu u_i^{(0)}\Lambda    + \nu v^{(0)}_i\Lambda^2   
= -2i\Lambda (\mu \sin\hat\theta_i + \nu\Lambda \sin 2\hat\theta_i) ~,
\label{lacei}
\eeq
and the perturbative couplings of each monopole to its corresponding dual
vector field,
\beq
\frac{4\pi}{{g_D^2}_i} \simeq - \frac{1}{4\pi}
\log\left(\frac{|C_i^{(0)}|}{\Lambda^2_i}\right) ~,
\label{coup}
\eeq
show logarithmic variations among different $U(1)$ factors. \footnote{~In
other words,  we are dealing here with a macroscopic (classical) theory of the
Ginzburg--Landau type, and  we should consider the coupling  constant of the
$M_i$ and $\tilde{M}_i^{\dagger}$ {\em classical} fields to $V^D_i$: 
wave-particle duality connects $g_D$ with the running coupling constant of
the quantum theory through the formula $\hbar g_{Di} = g_{Di}(a^D_i\sim
|C^{(0)}_i|^{1/2} )$, the strong coupling limit becoming the classical limit
for the magnetically charged quanta \cite{Coleman}.}

Even in the close vicinity of the singularity, different magnetic $U(1)$
factors are coupled through $\tau_{ij}^{\rm off}$ \cite{Klemmetal,DS}. Exactly
at the singularity, \ie at $a^D_i = 0$, the generic expression proposed in
\cite{edemas} for these off-diagonal couplings is 
\beq
\tau_{ij}^{\rm off} = \frac{2i}{N^2\pi} \sum_{k=1}^{N-1} \sin k\hat\theta_i
\sin k\hat\theta_j \sum_{p,q=1}^N \tau_{pq}^{(0)} \cos k\theta_p
\cos k\theta_q ~,
\label{result}
\eeq
where $\tau_{pq}^{(0)}$ is given by
\beq
\tau_{pq}^{(0)} = \delta_{pq}
\sum_{k\neq{p}}\log(2\cos\theta_p-2\cos\theta_k)^2 - (1-\delta_{pq})
\log(2\cos\theta_p-2\cos\theta_q)^2 ~,
\label{logs}
\eeq
with $\theta_p = (p-1/2)\pi/N$ and $p,q=1,...,N$. In the case of $SU(3)$, for
example,
$\tau_{12}^{\rm off} = i/\pi \log{2}$ \cite{Klemmetal,DS,edemas}. These
interactions are also present in the effective potential obtained from the
terms in square brackets of (\ref{efflag}),
\begin{eqnarray}
V_{eff} & = & \frac{1}{2} b_{ij}^{-1}(\phi^D) (|\phi_{m_i}|^2 -
|\tilde{\phi}_{m_i}|^2) (|\phi_{m_j}|^2 - |\tilde{\phi}_{m_j}|^2) + 2
|\phi^D_i|^2 (|\phi_{m_i}|^2 + |\tilde{\phi}_{m_i}|^2) \nonumber \\
& + & b_{ij}^{-1}(\phi^D) (\sqrt{2} \phi_{m_i} \tilde{\phi}_{m_i} +
C_i(\phi^D)) (\sqrt{2} \phi_{m_j} \tilde{\phi}_{m_j} + C_j(\phi^D))^* ~.
\label{pottbose}
\end{eqnarray}
Notice that, $b_{ij}^{-1}$ being positive definite, the potential is either
positive or zero. Given the expectation values of the complex scalars,
\beq
\langle  \phi^D_i\rangle = a^D_i ~~~~~~~~~ \langle\phi_{m_i}\rangle = m_i 
~~~~~~~~~
\langle\tilde{\phi}_{m_i}\rangle = \tilde{m}_i ~,
\label{vevs}
\eeq
${\cal N}=1$ supersymmetric vacua are in one to one correspondence with 
zeroes of
$V_{eff}$:
\beqa
\sqrt{2} m_i \tilde{m}_i &\rule{0mm}{6mm}=& -C_i(a^D) ~, \label{minim1} \\
m_i a^D_i &\rule{0mm}{6mm}=& \tilde{m}_i a^D_i = 0 ~,~~~~~~~~~~~ 
      \label{minim2} \\
|m_i| &\rule{0mm}{6mm}=& |\tilde{m}_i| ~.
\label{minim3}
\eeqa
$ i=1,2,...\,N-1$. From (\ref{minim2}) we learn that monopole condensation  
can only occur  at hypersurfaces where $a^D_i = 0$ for some $i$. At the
maximal singularity, every $a^D_i$ vanishes, and it is clear from
(\ref{minim1})--(\ref{minim3}) that $N-1$ monopoles have a chance to 
condense.
While soft breaking is parametrized by $\mu$ and $\nu$, monopole
condensation is controlled by $C_i $. If  for some $j$ we have $a^D_j = 0$ 
 and adjust 
$C_j^{(0)} = 0$, the corresponding  $U(1)$ remains unbroken ($m_j =
\tilde  m_j  = 0$),
 and the vacuum is said to be ``partially
Higgsed''. Summarizing, the Higgs vacuum ${\cal H}$ at the maximal point is
given by
\beq
{\cal H} = \{m_i,\tilde{m}_i ~/~ |m_i|^2 = |\tilde{m}_i|^2
=|C_i^{(0)}|/\sqrt{2} ~,~~ \tilde m_j = -e^{i\beta_j^{(0)}} m^*_j\} ~,
\label{Higgsv}
\eeq
with   $C_i^{(0)} = |C^{(0)}_i |e^{i\beta^{(0)}_i}$ given in equation 
(\ref{lacei}). Since the absolute phases of $m_i$ are not fixed, it has the
topology of a torus of genus $g=N-1$. Equation (\ref{Higgsv}) shows that the
scalar components of  the monopole superfields condense in the vacua placed
at the maximal points. Although the presence of condensation suggests that
confinement indeed takes  place, some further analysis is required before
this can be definitively established. An important question to be answered is
whether the collimation of the electric (or dual magnetic) flux lines is
energetically favored or not. This is a dynamical issue that goes beyond the
simple vacuum analysis.

%%%%%%%%%%%%%%%%%%%%%%%%%%%%%%%%%%%%%%%%%%%%%%%%%%%%%%%%%%%%%%%%%%%%%%%%%%%%%%%
%%%%%%%%%%%                         Section III                    %%%%%%%%%%%%
%%%%%%%%%%%%%%%%%%%%%%%%%%%%%%%%%%%%%%%%%%%%%%%%%%%%%%%%%%%%%%%%%%%%%%%%%%%%%%%
\section{Bogomol'nyi Bound in the Ultrastrong Scaling Limit}

The resulting effective theory we have arrived at, in the bosonic sector, is
an Abelian $(N-1)$--Higgs model with coupled $U(1)$ factors and a quite
non-standard Higgs potential. The search for stable vortex solutions in the
complete system is a hard problem. On general grounds, one should not expect
to have BPS string solutions in spite of the fact that $\N=1$ supersymmetry
is enough, generically, to have BPS vortices in four dimensions \cite{HS,ENS}.
At least, this is the case of ${\cal N}=1$ QCD, where the strings are
conserved modulo $N$ so they cannot carry an additive conserved quantity such
as a central charge \cite{Witt97}. There is a limit, however, in which the
system simplifies and admits BPS vortices \cite{HSZ}. It happens whenever  
the condensation parameters (\ref{condens}) are independent of $\phi^D$,
something that corresponds to linear perturbations in the superpotential
(\ref{superp}), \ie Fayet--Iliopoulos terms. This kind of terms together with
properly normalized quartic potentials are known to lead to Abelian
Higgs models that admit BPS vortices \cite{ENS,EN,ENS2}. Taking into
account (\ref{ui})--(\ref{vi}), one should consider $\Lambda\to\infty$ and
small values of the soft breaking parameters $\mu\to 0$ and $\nu\to 0$, such
that
$\mu\Lambda$ and 
$\nu\Lambda^2$ remain finite. In this ``ultrastrong'' limit,  
$C_i(\phi_D)\to C_i^{(0)}$ are constants, and one can easily check that
setting
$a^D_i = \langle\phi^D_i\rangle = 0$ is a consistent constraint.  One may
then study the existence of extended solutions in the remaining fields.

The (bosonic part of the) effective Lagrangian adopts the following form:
\begin{eqnarray}
{\cal L}_{eff,~B}^{{\cal N}=1} & = & - \frac{1}{4} b_{ij}^{(0)} 
 F_{\mu\nu}^i  F^{j\,\mu\nu}  + (D_\mu\phi_{m_i})^* D^\mu\phi_{m_i} +
D_\mu\tilde{\phi}_{m_i} (D^\mu\tilde{\phi}_{m_i})^* \nonumber \\
& - & \left[ \frac{1}{2} b_{ij}^{(0)} D_i^{(0)} D_j^{(0)} + b_{ij}^{(0)}
{F_i^{(0)}}^* F_j^{(0)}
\right] ~,
\label{efflagdos}
\end{eqnarray}
where $b_{ij}^{(0)}$ stands for the actual value of $b_{ij}$ at the maximal
singularity and $D_i^{(0)}, ~F_i^{(0)}$ are obtained from (\ref{DyF}) by
replacing $b_{ij}$ with $b_{ij}^{(0)}$. It is now feasible to give an
expression {\em \`a la Bogomol'nyi} \cite{Bogo} for the energy per unit
length corresponding to static and magnetically neutral ($(A^D_0)_i = 0$)
vortex-like configurations (\ie configurations with translational symmetry
along one axis) by means of the remainder ${\cal N}=1$ supersymmetry
\cite{HS,ENS} (see also \cite{EN} where a multi Higgs system has been
treated). Indeed, the energy density can be rearranged as follows:
\begin{eqnarray}
{\cal E}_{eff} &=& \frac{1}{2} b_{ij}^{(0)} \left( F_{12}^i \pm D_i^{(0)}
\right) \left( F_{12}^j \pm D_j^{(0)} \right) + \left| (D_1 \pm i D_2)
\phi_{m_i} \right|^2 + \left| (D_1 \pm i D_2) \tilde{\phi}_{m_i} \right|^2
\nonumber \\
&+& b_{ij}^{(0)} F_i^{(0)} {F_j^{(0)}}^* \mp \epsilon_{ab} \partial_a {\cal
J}_b ~,
\label{bogouno}
\end{eqnarray}
where 
the last term, corresponding to the current ${\cal J}_b =
-i(\phi^*_{m_i} D_b\phi_{m_i} + \tilde{\phi}^*_{m_i} D_b\tilde{\phi}_{m_i})$,
does not contribute to the string tension for finite energy configurations. It
is easier to analyze this system in a different set of variables,  obtained
from the above ones by means of an $SU(2)_R$ transformation yielding  
\begin{eqnarray}
{D}_i^{(0)} & ~~~ \longrightarrow ~~~ & \hat{D}_i^{(0)} = -  \sqrt{2}
~\mbox{\rm Re}(e^{i\alpha}F_i^{(0)}) ~, \label{trans1} \\
\rule{0mm}{8mm} \sqrt{2} {F}_i^{(0)} & ~~~ \longrightarrow ~~~ & \sqrt{2}
\hat{F}_i^{(0)} = - e^{-i\alpha} \left( D_i^{(0)} + i\sqrt{2} ~\mbox{\rm
Im}(e^{i\alpha}F_i^{(0)}) \right) ~, \label{trans2} \\
\phi_{m_i} & ~~~ \longrightarrow ~~~ & \hat\phi_{m_i} = - \frac{i}{\sqrt{2}}
(\phi_{m_i} - e^{-i\alpha} \tilde{\phi}^*_{m_i}) ~, \label{trans3} \\
{\tilde{\phi}}_{m_i}^* & ~~~ \longrightarrow ~~~ & \hat{\tilde{\phi}}_{m_i}^*
= - \frac{i}{\sqrt{2}} e^{i\alpha} (\phi_{m_i} + e^{-i\alpha}
\tilde{\phi}^*_{m_i}) ~.
\label{trans4}
\end{eqnarray}
The tension $\sigma_{eff} = \int d^2x ~{\cal E}_{eff}$ now reads
\begin{eqnarray}
\sigma_{eff} & = & \int d^2x \left[ \frac{1}{2} b_{ij}^{(0)} \left( F_{12}^i
\pm \hat{D}_i^{(0)} \right) \left( F_{12}^j \pm \hat{D}_j^{(0)} \right) +
\left| (D_1 \pm i D_2) \hat\phi_{m_i} \right|^2 \right. \nonumber \\
& + & \left. \left| (D_1 \pm i D_2) \hat{\tilde{\phi}}_{m_i} \right|^2 +
b_{ij}^{(0)} \hat{F}_i^{(0)} \hat{F}_j^{(0) ~*} \mp \sqrt{2} F_{12}^i
~\mbox{\rm Re}(e^{i\alpha}C_i^{(0)}) \right] ~.
\label{bogodos}
\end{eqnarray}
The last term breaks explicitely $SU(2)_R$ symmetry. 
Finiteness of the string tension demands regularity of the fields on
${\mathbb R}^2$, and vanishing of the potential energy, field strenghts and
covariant derivatives at infinity. Altogether, these requirements make the
space of solutions to split into ${\mathbb Z}^{N-1}$ disconnected pieces that
differ by the winding numbers of each $\phi_{m_i}$ over the border of the
plane. The electric-fluxes label these sectors. In particular, in the
$(n_1,n_2,...,n_{N-1})$--sector they are
\beq
\Phi_j=-\int d^2x ~F_{12}^j =   2\pi n_j   ~, ~~~~~~~~
j=1,2,...,N-1 ~.
\label{fluxes}
\eeq

The string tension of possible vortex configurations with topologically
quantized $(n_1,n_2)$ electric flux, exhibits a Bogomol'nyi bound
\beq
\sigma_{eff} \geq  4\sqrt{2}\pi\Lambda  \sum_i
|(\mu~\sin\hat\theta_i + \nu\Lambda~\sin 2\hat\theta_i)
\cos(\alpha+\beta_i^{(0)}) ~n_i| ~,
\label{bogobound}
\eeq
which is saturated for configurations solving the following set of first order
equations:
\begin{eqnarray}
F_{12}^i = \pm \sqrt{2} ~\mbox{\rm Re}(e^{i\alpha}F_i^{(0)}) ~, &
~~~~~~~~ & D_i^{(0)} + i\sqrt{2}~\mbox{\rm Im}(e^{i\alpha}F_i^{(0)}) = 0 ~,
\label{boguno} \\ (D_1 \pm i D_2) \hat\phi_{m_i} = 0 ~, & ~~~~~~~~ & (D_1 \pm
i D_2) \hat{\tilde{\phi}}_{m_i} = 0 ~. \label{bogdos}
\end{eqnarray}
The second equation in (\ref{boguno}) implies
\beq
|\phi_{m_j}| = |\tilde{\phi}_{m_j}| ~~~~~~~~~~~ \mbox{\rm
Im}\left(e^{i\alpha} \left[ \sqrt{2} \phi_{m_j} \tilde{\phi}_{m_j} +
C_j^{(0)} \right] \right) = 0 ~.
\label{constrs}
\eeq
These constraints should hold at any point, in particular, at zeroes of the
Higgs field. Thus 
\beq
\phi_{m_i} = - e^{i\beta_i^{(0)}} \tilde{\phi}_{m_i}^* ~.
\label{constr}
\eeq
with $\alpha + \beta_i^{(0)} = 0$ or $\pi$. Consequently,  for $i\neq j$,
$\beta_{ij}^{(0)} \equiv \beta_i^{(0)} - \beta_j^{(0)} = 0$ or $\pi$.
Summarizing, {\em there are no BPS electric vortices in the system unless the
complex numbers $C_i^{(0)}$ are aligned or anti--aligned}. This alignment, in
turn, requires supersymmetry breaking parameters to have no relative complex
phases. Notice that this corresponds to having a CP invariant bare Lagrangian.
For definiteness, in the case of $SU(3)$ one easily sees that
\beq
C_1^{(0)} = \sqrt 3 \Lambda (\mu + \nu\Lambda) ~~~~~~~ ; ~~~~~~~
C_2^{(0)} = \sqrt 3 \Lambda (\mu - \nu\Lambda) ~, \label{cesu}
\eeq
so that $\beta_{21}^{(0)}= 0$ or $\pi$  if and only if arg$(\nu\Lambda) =$
arg$\mu + n\pi$ and $|\nu\Lambda|<|\mu|$ or $|\nu\Lambda|>|\mu|$ respectively.

A comment is in order at this point regarding the string tensions of unit
vortices, whose existence will be discussed below. It is immediate to read,
from (\ref{bogobound}), the string tension of electric vortices carrying
a single flux quantum $n_k = 1, ~n_{i\neq k} = 0$. Up to a common 
factor, it is given by
\beq
T_k \propto \Lambda f_N(k) ~~~~~~~~~~~ f_N(k) = |\mu \,\sin\hat\theta_k +
\nu\Lambda \,\sin 2\hat\theta_k| ~.
\label{tensions}
\eeq
This  result   makes clear the dependence of $f_N(k)$
on the  supersymmetry breaking deformation entering the superpotential.
It generalizes previous results in \cite{DS,HSZ,Str} and, in particular,
  it shows that for
perturbations other than the quadratic one,  the string tensions are modified
with respect to those in the above mentioned results. In particular, notice
that when
$\mu$ and $\nu$ do not vanish it is possible to have different string 
tensions even in the case of $SU(3)$ and, in general, $T_k \neq T_{N-k}$.  

%%%%%%%%%%%%%%%%%%%%%%%%%%%%%%%%%%%%%%%%%%%%%%%%%%%%%%%%%%%%%%%%%%%%%%%%%%%%%%%
%%%%%%%%%%%                         Section IV                    %%%%%%%%%%%%%
%%%%%%%%%%%%%%%%%%%%%%%%%%%%%%%%%%%%%%%%%%%%%%%%%%%%%%%%%%%%%%%%%%%%%%%%%%%%%%%
\section{Aligned vacua: critical vortices}

We will focus hereafter on the case of $SU(3)$. When the constants $\mu$ and
$\nu$ are fine tuned in such a way that the phases of the two complex energy
scales $C_1^{(0)}$ and $C_2^{(0)}$ are either aligned or antialigned, \ie
$\beta _{21}^{(0)}  = 0$ or $\pi$ respectively, we are at the self dual
point. The Bogomol'nyi equations (\ref{boguno})--(\ref{bogdos}), after
(\ref{constr}), read
\begin{eqnarray}
F_{12}^i & = & \pm \frac{1}{2} b_{ij}^{(0)\,-1} \epsilon_j (|\varphi_j|^2
- v_j^2) ~, \label{buno} \\
(D_1 & \pm & i \epsilon_j D_2) \varphi_j = 0 ~, \label{bdos}
\end{eqnarray}
where  $\epsilon_j = e^{i(\alpha+\beta_j^{(0)})} = \pm 1$ and $b_{ij}^{(0)}$
is
\beq
b_{ij}^{(0)} = \left( 
\begin{array}{cc} g_{D,1}^{-2} & \frac{1}{4\pi^2}\log{2} \\
\frac{1}{4\pi^2}\log{2} & g_{D,2}^{-2} \end{array} 
\right) ~,
\label{coupling}
\eeq
 Also, we have
performed, for convenience, some redefinitions of the fields, $\varphi_j = 2
\phi_{m_j}$, and parameters, $v_j^2 = 2\sqrt{2} |C_j^{(0)}|$. Let us further
remark that Eq.(\ref{buno}) gives an unusual contribution to the electric
field of each dual $U(1)$ factor from zeroes of both Higgs fields. This is a
straight consequence of the presence of off-diagonal couplings and leads to
interesting results. It is clear that solutions to (\ref{buno})--(\ref{bdos})
also satisfy the Euler-Lagrange equations. Without loss of generality, we can
adjust $\alpha$ so that $\epsilon_1 = +1, \epsilon_2 \equiv \epsilon =
e^{i\beta _{21}^{(0)}} = \pm 1$. Let us focus on the BPS solutions with upper
sign. The first order system can be written as
\begin{eqnarray}
F_{12}^1 & = & ~~\lambda_1 (|\varphi_1|^2-v_1^2) - \epsilon \gamma\,
(|\varphi_2|^2-v_2^2) ~, \label{eq1} \\
F_{12}^2 & = & -\gamma \,(|\varphi_1|^2-v_1^2) + \epsilon \lambda_2 
(|\varphi_2|^2-v_2^2) ~, \label{eq2} \\
(D_1 & + &\, i D_2)\, \varphi_1=0 ~, \label{eq3} \\
(D_1 & + & i\epsilon D_2) \varphi_2 = 0 ~, \label{eq4}
\end{eqnarray}
with
\beqa
\lambda_i &=&b_{ii}^{(0)\,-1}=  \left(1-\frac{g_{D,1}^2g_{D,2}^2}{16\pi^2}
\log^2{2}\right)^{-1} ~\frac{g_{D,i}^2}{2} \label{lyg1}\\
 \gamma &=&
b_{12}^{(0)\,-1} = \frac{\log{2}}{8\pi^2} 
 (g_{D,1}^2\lambda_2 + g_{D,2}^2\lambda_1)~.
\label{lyg2}
\eeqa
Note that, as we are in the weak $g_D$-coupling regime, $\gamma < \lambda_i$.
Naively one would suspect that in the scaling limit we are interested, the
system diagonalizes. Notice however the important fact that the
relative factor between $\lambda_i$ and $\gamma$ vanishes only logarithmically.
Hence, for example, setting $|C_i|/\Lambda_i^2 \sim 10^{-10}$ in (\ref{coup})
yields $\gamma\sim (\log  2/5)  \lambda_i \sim 0.13 \,\lambda_i$.

The topology of the configuration space determines global properties of the
solutions in two ways: the quantization of the fluxes is due either to the
asymptotics of the $A_j$ fields or to the existence of a prescribed number of
zeroes of the $\varphi_j$. These global inputs should be made compatible with
the differential equations, as it happens in the Abelian Higgs model. In the
present situation things are less clear; from Eqs.(\ref{eq3})--(\ref{eq4}),
 where no mixing between both $U(1)$s shows up,
one reads the electric fluxes using Stokes theorem and the asymptotics of
$A_j$. On the other hand,
Eqs.(\ref{eq1})--(\ref{eq2}) mix the factors and both $\varphi_1$ and
$\varphi_2$ contribute together to each $F_{12}^i$. In this respect, our
system is quite awkward as compared with other non-diagonal models as, for
example, non-relativistic non-abelian Chern-Simons theories \cite{Dunne}, in
which the same mixing appears in the field strength and covariant derivative
equations. Here, there is mixing in the former but not in the latter, and
given such an asymmetry, it is much more difficult to show whether the local
equations and the global conditions reconcile or not.

On general grounds, it is reasonable to expect that  the equations
(\ref{eq1})--(\ref{eq4}) will  exhibit solutions in the topological sector
$(n_1,n_2)$ with $n_1,n_2$ representing the integrated flux of  an
``ensemble'' of noninteracting vortices located at different (maybe
coincident) positions. Indeed, the smallness of the ratio $\gamma/\lambda_i$
suggests to   consider this system as a pertubation of the diagonal
situation, so that the above solutions would come out from continuous
deformations of the standard critical Abrikosov vortices. Only in some
simple cases, the question about the existence of solutions can be answered by
taking advantage of known results from  the standard Abelian Higgs model. This
will be done in the following two situations

\begin{description}
\item { $\bullet$ \bf Solutions of type $(n,0)$ and $(0,n)$}.

Clearly it will be enough to prove existence of one type, say $(n,0)$.
Assume  therefore that $\varphi_2 = |\varphi_2|e^{i\xi_2}$ is nowhere
vanishing on the finite transverse plane. As usual,  (\ref{eq4}) couples
$\xi_2$ and $A_2$. So, if $|\varphi_2|$ has nowhere a zero,  regularity of
the phase  enforces $A_2$ to have vanishing circulation around any  loop. By
Stokes theorem $F_{12}^2 = 0$ everywhere, an inserting this back into
(\ref{eq2}) yields a constraint that correlates the profiles of
$|\varphi_1|$  and $|\varphi_2|$,
\beq
|\varphi_2|^2  = \epsilon\frac{\gamma}{\lambda_2} (|\varphi_1|^2-v_1^2) +
v_2^2 ~.
\label{ddeun}
\eeq
Existence of the required vortex profile for $|\varphi_1|$ can be proved by
inserting (\ref{ddeun}) into $(\ref{eq1})$, which leads to the standard
Bogomol'nyi equations for the critical Abelian Higgs model (after a suitable
re-normalization of the Higgs field)
\begin{eqnarray}
F_{12}^1 &=& \lambda_1\left(1-\frac{\gamma^2}{\lambda_2}\right)
(|\varphi_1|^2 - v_1^2) ~, \label{iuy} \\
(D_1  &+&  i D_2)\varphi_1~ = ~0 \label{iuy2} ~.
\end{eqnarray}
We learn from (\ref{ddeun}) that if $|\varphi_1|^2$ ranges from $0$ (at the
origin)  up to $v_1^2 $ (at infinity),
$|\varphi_2|^2$ will correspondingly interpolate between
$ - \epsilon\frac{\gamma}{\lambda_2}  v_1^2  + v_2^2$
and $v_2^2$.
To remain consistent with our initial asumption that $|\varphi_2|$ vanished
nowhere we must set  either  $\beta_{21}^{(0)} = 0$  
with 
$v_2^2 > 
\frac{\gamma}{\lambda_2}  v_1^2$,  or else $ \epsilon = -1$, \ie,
$\beta_{21}^{(0)} = \pi$. We observe that the latter possibility is less
contrived.

\item { $\bullet$ \bf Solutions of type $(n,n)$ for a single perturbation}.

Let us briefly consider the case of $SU(3)$ ${\cal N}=2$ supersymmetric
Yang--Mills theory softly broken to ${\cal N}=1$ only by means of a single
Casimir operator, \ie $\mu = 0$ or $\nu = 0$. In both cases, $\beta_{21}^{(0)}
= 0$ or $\pi$, and the theory is critical. Moreover, $\lambda_1 =
\lambda_2 \equiv \lambda$, $C_1^{(0)} = C_2^{(0)}$ and hence $v_1=v_2\equiv
v$, so that the Bogomol'nyi equations have an almost trivial solution of
vorticity $(n,n)$ (or $(n,-n)$), by imposing the ansatz $\varphi_j \equiv
\varphi$, $~A_j \equiv A$  (or $\varphi_2^* = \varphi_1 \equiv\varphi$,
$~-A_2 =  A_1\equiv A$) in the case  $\beta_{21}^{(0)} = 0$ (or $\pi$). The
system is again reduced, after a suitable normalization of the Higgs field,
to the critical Abelian Higgs model
\beqa
F_{12}  &=& \pm(\lambda - \epsilon \gamma)
(|\varphi |^2 - v^2) ~, \label{iuyn} \\
(D_1  &\pm&  i D_2)\varphi ~ = ~0 \label{iuy2n} ~, ~~~~~~~~
\epsilon = e^{i\beta_{21}^{(0)}} ~.
\end{eqnarray}
It is crucial, for the system to admit regular solutions, that $\gamma
<\lambda$ as it indeed happens. As it is well known, the general solution to
this sytem represents an assembly of $n$ separated vortices centered at the
zeroes of $\varphi$. In our case, every such zero is doubled and we have
assemblies of $n$ couples of superimposed vortices of both $U(1)$ fields.

~~Also, self-dual configurations in which the center of the vortices of
different types split apart, can be easily constructed along the lines in
\cite{Wein,Ruback}. To see this, we perturb one of the solutions just
described for
$\beta_{21}^{(0)}= 0$
\beq
\varphi_j^\prime=\varphi_j+\delta\varphi_j ~, ~~~~~~~~~~~~   
A_j^\prime=A_j+\delta A_j ~,
\eeq
and linearize the self-duality equations to get
\begin{eqnarray}
-4i\partial_z\delta A_1 - 2\lambda\varphi^*\delta\varphi_1 +
2\gamma\varphi^*\delta\varphi_2 & = & 0 ~, \label{6a} \\  
-4i\partial_z\delta A_2 + 2\gamma\varphi^*\delta\varphi_1 -
2\lambda\varphi^*\delta\varphi_2 & = & 0 ~, \label{6b} \\ 
ig_D\varphi\delta A_j + (\partial_{\bar{z}}+ig_DA_j)\delta\varphi_j & = & 0 ~,
\label{lin3}
\end{eqnarray}
where we use the notation $\partial_z=\frac{1}{2}(\partial_1-i\partial_2)$,
$A_j=\frac{1}{2}[(A_1)_j+i(A_2)_j], ~j=1,2$, and fix the gauge conditions as
\begin{eqnarray}
\partial_c(\delta A_c)_1 & = & -\lambda|\varphi|^2\delta\Omega_1 +
\gamma|\varphi|^2\delta\Omega_2 ~, \\
\partial_c(\delta A_c)_2 & = & \gamma|\varphi|^2\delta\Omega_1 -
\lambda|\varphi|^2\delta\Omega_2 ~.
\end{eqnarray}
By writing $\delta \varphi_j=\varphi\xi_j$ and using (\ref{lin3}), the vector
perturbations are found to be $\delta
A_j=\frac{i}{g_D}\partial_{\bar{z}}\xi_j$ and the system of linearized
equations reduces to
\beq
\nabla^2 W_\pm=2(\lambda\mp\gamma)g_D|\varphi|^2W_\pm ~,
\label{59}
\eeq
with $W_\pm=\xi_1\pm\xi_2$. Notice that in both equations
$(\lambda\mp\gamma)g_D>0$. Although they have not regular square-integrable
solutions, we can admit singular ones provided the singularities of $\xi_j$
fit with the zeroes of $\varphi$ in such a way that $\delta\varphi_j$ is
well-behaved. Take for instance the case of a radially symmetric solution of
vorticity $n$ centered at the origin of the complex plane. Then, for small $z$
\beq
\varphi(z,\bar{z})\simeq z^n ~,
\eeq
and a singularity of $W_\pm$ at the origin is harmless if its order is lower
or equal than $n$. Equation (\ref{59}) has indeed solutions with such a
behaviour \cite{sa}. To be exact, two sets of linearly independent self-dual
perturbations $W_\pm^m(z,\bar{z}),m=1,2,3,.....,n$ with
\beq
W_\pm^m(z,\bar{z})\simeq z^{-m},\ \ \ \ \ \ \ z\simeq 0 ~.
\eeq
In particular, if we consider $W_\pm=-aW_\pm^m$, we get, near the origin,
\beq
\xi_1\simeq -az^{-m} ~~~~~~~~~~~~~~~ \xi_2\simeq 0 ~,
\eeq
so that
\beq
\varphi_1^\prime \simeq z^{n-m}(z^m-a) ~~~~~~~~~~~~~~~~
\varphi_2^\prime \simeq z^n ~.
\eeq
This perturbation realizes the splitting of a $(n,n)$ vortex at the origin
into a $(n-m,n)$ at that point and  $m$ $(1,0)$ vortices located at the $m$
roots of the coefficient $a$. The analysis for $\beta_{21}^{(0)} =
\pi$ (\ie $\epsilon = -1$) is totally equivalent and yields nothing
but vortices of type 1 and anti-vortices of type 2 or viceversa,  moving  
freely with respect to each other.
\end{description}

For the  general analysis, following Jaffe and Taubes \cite{JT}, the Higgs
fields should be ``couched'' as 
\beq
\varphi_j \equiv v_j e^{\frac{1}{2}(u_j+i\Omega_j)} ~,
\eeq
to recast the Higgs system in the following form
\begin{eqnarray}
\nabla^2 u_1 & = &~~ 2\lambda_1 v_1^2 (e^{u_1}-1) - 2 \epsilon\gamma
v_2^2 (e^{u_2}-1) + \varepsilon_{bc}\partial_b\partial_c\Omega_1 ~,
\label{jt1} \\ \nabla^2 u_2 & = & -2\gamma v_1^2 (e^{u_1}-1) +
2\epsilon\lambda_2  v_2^2 (e^{u_2}-1) +
\varepsilon_{bc}\partial_b\partial_c\Omega_2 ~.
\label{jt2}
\end{eqnarray}
The gauge fields are determined by
\begin{eqnarray}
(A_c)_1 & = & - \frac{1}{2 } (\partial_c\Omega_1 +
\varepsilon_{ca}\partial_a u_1) ~, \label{jt3} \\ (A_c)_2 & = & -
\frac{\epsilon}{2 } (\partial_c\Omega_2 +
\varepsilon_{ca}\partial_a u_2) ~. \label{jt4}
\end{eqnarray}
At each $(n_1,n_2)$ sector, regularity implies that $\varphi_j$ has exactly
$n_j$ zeroes on $\mathbb{C}$, say $z_1^j,z_2^j,\ldots,z_{n_j}^j$. Also, these
are the only points at which the singularities of the phases can occur. We can
then choose the particular gauge
\beq
\Omega_j(z,\bar{z}) = 2\sum_{l=1}^{n_j}\arg (z-z_l^j) ~,
\eeq
in which the problem reduces to
\begin{eqnarray}
\nabla^2 u_1 & = &~~ 2\lambda_1 v_1^2 (e^{u_1}-1) - 2\epsilon\gamma 
v_2^2 (e^{u_2}-1) + 4\pi\sum_{l=1}^{n_1}\delta(z-z_l^1) ~, \label{redto1} \\
\nabla^2 u_2 & = & -2\gamma v_1^2 (e^{u_1}-1) + 2 \epsilon\lambda_2
v_2^2 (e^{u_2}-1) + 4\pi\sum_{l=1}^{n_2}\delta(z-z_l^2) ~, \label{redto2}
\end{eqnarray}
where both $u_j$ should vanish at space infinity. The general analysis is  
involved, and usually goes through by numerical relaxation techniques or hard
Sovolev estimates.

%%%%%%%%%%%%%%%%%%%%%%%%%%%%%%%%%%%%%%%%%%%%%%%%%%%%%%%%%%%%%%%%%%%%%%%%%%%%%%%
%%%%%%%%%%%                         Section V                     %%%%%%%%%%%%%
%%%%%%%%%%%%%%%%%%%%%%%%%%%%%%%%%%%%%%%%%%%%%%%%%%%%%%%%%%%%%%%%%%%%%%%%%%%%%%%
\section{Hybrid type II vortices}

By itself, the  Abelian Higgs model we are dealing with is worth a detailed
analysis. For the moment, and awaiting a sounder analytical or numerical
study of  its solutions, aside from the  two simplified samples considered
above  little can be said about the generic $(n_1,n_2)$ vortex solution. An
interesting peculiarity comes from the fact that there are only two overall
choices of signs available in equations (\ref{buno}) and (\ref{bdos}): either
upper or lower sign have to be  taken simultaneously on all the equations or,
else, the bound (\ref{bogobound}) will not be saturated. This should be
contrasted with the situation in the standard diagonal Abelian Higgs model,
where each $U(1)$ can be  conjugated independently. To  better grasp what is
going  on let us consider the Bogomol'nyi equations (\ref{eq1})-(\ref{eq4})
with $\beta_{21}^{(0)} = 0$
\begin{eqnarray}
F_{12}^1  & = &  \pm(\lambda_1 W_1 -   \gamma  W_2)~, \label{eq11} \\
(D_1 & \pm &  i D_2)\, \varphi_1=0 ~, \label{eq33} \\
F_{12}^2  & =\rule{0mm}{7mm} &    \pm   (\lambda_2 W_2 - \gamma \,W_1) 
 ~, \label{eq22} \\
(D_1 & \pm & i D_2) \varphi_2 = 0 ~, \label{eq44}
\end{eqnarray}
with $W_i = (|\varphi_i|^2 - v_i^2)$\,. If $\gamma<<\lambda_1,\lambda_2$,
$(\pm n_1,\pm n_2)$ vortex with $n_1,n_2>0$ come from solutions to the
previous equations with the upper (lower) sign which should correspond to
deformations of analogous  configurations in the case $\gamma = 0$. In the
diagonal limit $\gamma= 0$ the vortex-antivortex solutions $(\pm n_1,\mp
n_2)$ would also solve the previous equations but with a choice of sign for
(\ref{eq11})--(\ref{eq33}) and the opposite one for
(\ref{eq22})--(\ref{eq44}). If $\gamma\neq 0$, as is now the case, solutions
with this second choice of sign do not saturate the bound  (\ref{bogobound})
and, indeed, there is an energy remnant coming from the off-diagonal piece
$
{\cal E} = \pi \, |n_1 v_1^2 + n_2 v_2^2| + \delta{\cal E} $
\beq 
\delta{\cal E} = \int d^2x\,  \delta \sigma_{eff} = 
\int d^2x\, 2 b_{12}^{(0)} F_{12}^1 F_{12}^2 =
\frac{\log{2}}{2\pi^2}\int d^2x\, F_{12}^1 F_{12}^2 \label{overlap}
\eeq
For anti-aligned magnetic fields, this extra term is negative and
tends to increase the overlap by attracting the  cores of vortices of
different kind.  

A similar reasoning can be carried out of $\beta_{21}^{(0)} = \pi$.
In this case, the equations read 
\begin{eqnarray}
F_{12}^1 & = &  \pm(\lambda_1 W_1 +   \gamma  W_2)~, \label{eq111} \\
(D_1 & \pm & i D_2)\, \varphi_1=0 ~, \label{eq333} \\
F_{12}^2 & =\rule{0mm}{7mm} &    \mp   (\lambda_2 W_2 + \gamma \,W_1) 
 ~, \label{eq222} \\
(D_1 & \mp & i D_2) \varphi_2 = 0 ~, \label{eq444}
\end{eqnarray}
and critical configurations are naturally of the form
$(\pm n_1,\mp n_2),~~n_1,n_2\geq 0$ saturating the bound ${\cal E} = \pi
\,|n_1 v_1^2 - n_2 v_2^2\,|$. Here, in contrast, vortex-vortex solutions of
the form  $(\pm n_1,\pm n_2)$ would lead to the same 
energy surplus as in (\ref{overlap}). But now $\delta {\cal E}\geq 0 $
for aligned magnetic fields, and this term decreases by minimizing the
overlap, hence by taking the cores far appart. 

In summary, to a first approximation, we see that, if not neutral,
vortex-vortex (vortex-antivortex) configurations behave repulsively
(attractively) as in type II superconductors. Since this interaction involves
vortices of different $U(1)$'s, we speak of an ``hybrid type II'' phase.

\vskip0.3cm

Let us discuss the peculiarities that arise whenever one tries to model
confinement in  the present scenario. First we fix some notation for
convenience: the chromoelectric fluxes $(n_1,n_2)$ of the basic vortices
arising in the dual Meissner effect are $(1,0)$ (``vortex 1'') and $(0,1)$
(``vortex 2''). In turn, quarks enter the system as external probes with
chromoelectric charges
$(Q_1,Q_2)$ equal to $(1,0)$ (``red quark''), $(0,-1)$ (``blue quark'') and
$(-1,1)$ (``yellow quark''). $(h_1,h_2)$ is the ``monopole'' basis of the
Cartan algebra of the dual $S\check{U}(3)$ group and the fundamental BPS
monopoles correspond to the simple co-roots of $SU(3)$. In other words, the
chromomagnetic charges of the $\varphi_i$--field quanta is $h_i=1, h_{j\neq
i}=0$. 
Consider now, for example, the case $\beta_{21}^{(0)} =0$. According to
our previous analysis, chromoelectric flux tubes of both $(1,0)$ and $(0,1)$
type form in response to parallel external electric fields $\vec{E}_1$ and
$\vec{E}_2$. Vortices of type 1 end at pairs of red quark-antiquark and
vortices of type 2 finish at pairs of blue antiquark-quark. There is therefore
confinement of red and blue quarks in a critical phase between Type I and Type
II superconductivity, whereas the yellow quark confinement occurs in a hybrid
Type II phase. The weak repulsion between the vortex 1/antivortex 2 pair pull
slightly apart the flux lines from each other. Thus, the quark/antiquark
potential energy would  increase slower than linearly with the
distance, and one is allowed to expect deviations from the area law, but the
force is still confining.  If, instead, $\beta_{21}^{(0)}=\pi$, a pair
of yellow quark-antiquark will now be joined by a stable and non-interacting
vortex 1/antivortex 2 pair of flux tubes. In conclusion, the cases
$\beta_{21}^{(0)} =0$ or $\pi$ can be physically distinguished by the
behaviour of the yellow quark-antiquark force.
 At large separation W-pair production leads to instability of the string and
the lowest string tension governs the large distance regime
\cite{DS,HSZ}.

\vskip0.3cm

In the framework of condensed matter it is well known the fact that, in 
standard type II superconductivity on a finite piece of material, though 
mutually repelling, vortices tend to form a regular pattern by lying at the
sites of a triangular lattice. This fact can be reproduced analytically by
variational methods \cite{abr}. We expect a similar situation here, the
difference being that now repulsion involves vortex cores of distinct Higgs
fields. Upon substitution of (\ref{constr}) into (\ref{efflagdos}), the exact
second order equations with $\beta_{21}^{(0)}=\pi$, corresponding to vortices
of type 1 and 2, in a finite piece of material
\begin{eqnarray}
b^{(0)}_{11}\partial_a(F_{ab})_1 + b^{(0)}_{12}\partial_a(F_{ab})_2 & = &
\frac{i }{2} (\varphi_1^*D_b\varphi_1 - \varphi_1D_b\varphi_1^*) ~, \\
b^{(0)}_{22}\partial_a(F_{ab})_2 + b^{(0)}_{21}\partial_a(F_{ab})_1 & = &
\frac{i }{2} (\varphi_2^*D_b\varphi_2 - \varphi_2D_b\varphi_2^*) ~, \\
D_cD_c\varphi_1 & = & -\frac{1}{\sqrt{2}}\varphi_1^*b^{(0)-1}_{1j}
(|\varphi_j|^2-v_j^2) (-1)^{\beta_{1j}} ~, \\
D_cD_c\varphi_2 & = & -\frac{1}{\sqrt{2}}\varphi_2^*b^{(0)-1}_{2j}
(|\varphi_j|^2-v_j^2) (-1)^{\beta_{2j}} ~,
\end{eqnarray}
should now be supplemented with periodic boundary conditions. Thus, the system
of differential equations is defined in a torus of modular parameter $\tau =
L_2/L_1 ~e^{i\theta}$. We have chosen the $x_1$-axis as the direction of the
first $L_1$ periodicity; the length and direction of the second periodicity is
determined by $L_2e^{i\theta}$. Application of the Rayleigh-Ritz variational
method as in
\cite{abr} plus previous work on the r\^{o}le of Riemann Theta functions in
magnetic systems
\cite{matmar}, suggest the field configurations
\begin{eqnarray}
\varphi_1&=&\sum_{m_1\in{\bf Z}}C_{m_1}\exp[in_1m_1{\rm Im}z -\frac{1}{2}({\rm
Re}z-n_1m_1)^2]~,\\
\varphi_2&=&\sum_{m_2\in{\bf Z}}C_{m_2}\exp[in_2m_2{\rm Im}z -\frac{1}{2}({\rm
Re}z-n_2m_2)^2]~,
\end{eqnarray} 
where $n_1$, $n_2$ are integers and
${\displaystyle
z=\sqrt{g_D(\lambda-\gamma)}~\left(\frac{x_1+ix_2}{L_1}\right)}$, as trial
functions to model extremals of the energy. In fact, the choice of the
coefficients $C_{m_1}$ and $C_{m_2}$ in such a way that
\begin{eqnarray}
\varphi_1^{n_1}(z) & = & \exp\{-\pi n_1\frac{({\rm Im}z)^2}{{\rm Im}\tau}\}
\prod_{l_1=1}^{n_1}\Theta \left[ 
\begin{array}{c} 0 \\ \frac{l_1}{n_1} \end{array}
\right](z|\frac{\tau}{n_1}) ~, \\
\varphi_2^{n_2}(z) & = & \exp\{-\pi n_2\frac{({\rm Im}z)^2}{{\rm Im}\tau}\}
\prod_{l_2=1}^{n_2}\Theta \left[ 
\begin{array}{c} \frac{1}{2} \\
\frac{l_2}{n_2}+\frac{1}{2}\end{array} 
\right](z|\frac{\tau}{n_2}) ~,
\end{eqnarray}
leads to (meta)-stable solutions to the field equations. Here $l_i=1,\ldots,
n_i$, and $\Theta [^a_b] (z|\tau)$ are the Riemann Theta functions with
characteristics, see \cite{matmar} and references quoted therein.

~

\vskip0.5cm

%%%%%%%%%%%%%%%%%%%%%%%%%%%%%%%%%%%%%%%%%%%%%%%%%%%%%%%
\begin{figure} 
\centerline{\psfig{figure=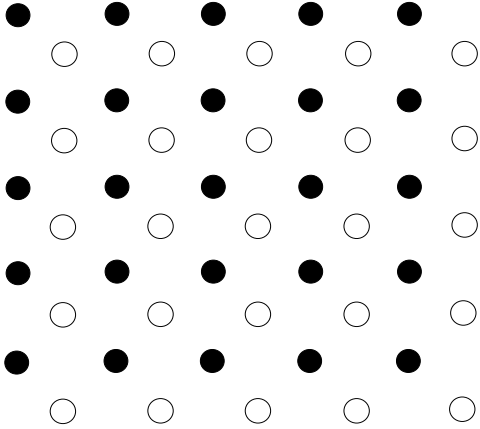,height=5.0cm,width=8cm}} 
\vspace{1.0cm} 
\caption{The Type II hybrid lattice.
Black and white circles represent the core of vortices corresponding to
different $U(1)^{'s}$.}
\end{figure}
%%%%%%%%%%%%%%%%%%%%%%%%%%%%%%%%%%%%%%%%%%%%%%%%%%%%%%%

Notice that the solution describes $n_1$ chromoelectric vortices, located
at the zeroes of $\varphi_1^{n_1}$, and $n_2$ vortices of the other kind
centered around the zeroes of $\varphi_2^{n_2}$. It corresponds therefore to
a hybrid static triangular lattice of vortices; see the Figure. One can check,
from a dynamical point of view that a configuration like this where a vortex
of type 1 is at the center of a square with vortices of type 2 at the vertices
and viceversa is stable against small fluctuations. 

%%%%%%%%%%%%%%%%%%%%%%%%%%%%%%%%%%%%%%%%%%%%%%%%%%%%%%%%%%%%%%%%%%%%%%%%%%%%%%%
%%%%%%%%%%%                         Section VI                    %%%%%%%%%%%%%
%%%%%%%%%%%%%%%%%%%%%%%%%%%%%%%%%%%%%%%%%%%%%%%%%%%%%%%%%%%%%%%%%%%%%%%%%%%%%%%
\section{Misaligned vacua}

As discussed earlier, there are no BPS vortices in the generic case where soft
breaking parameters are not aligned. We would however be interested in the
response of the BPS configuration when an infinitesimal misalignment
$\beta_{21}^{(0)}\equiv\varepsilon$  or 
$\beta_{21}^{(0)}\equiv\pi +\varepsilon$ is turned on. The dynamics of the
system drives the configuration off the constraint (\ref{constr}) which,
therefore, can no longer be imposed consistently. In fact, though the Higgs
mechanism yields a critical mass spectra for any value of
$\varepsilon$ (an obvious consequence of supersymmetry), the eigenvectors do
depend on this phase difference in such a way that when it is different from
$0$ or $\pi$, massive excitations do not respect the constraint
surface (\ref{constr}).

In the same vein as above, for small values of $\varepsilon$ we will
treat the system as a perturbation of the critical situation in which the net
effect of the misalignment  reflects itself in a  force between the former
noninteracting   vortices. The shortcut to obtain the sign of this force is
to split  the energy  (\ref{bogodos}) of the configuration  as a BPS
contribution plus an additional perturbation. Namely, after inserting the
ansatz   (\ref{constr}) into (\ref{efflagdos}), solutions to
(\ref{buno})--(\ref{bdos}) exhibit a string tension
$\sigma_{eff}=\sigma^{SD}_{ eff}+\delta\sigma_{eff}$ where 
$\sigma^{SD}_{eff}$ is given in (\ref{bogobound}) and
\beq 
\delta\sigma_{eff} = \,\epsilon\frac{
\gamma\varepsilon^2}{8} \int d^2x\, (|\varphi_1|^2 - v_1^2)  (|\varphi_2|^2 -
v_2^2)   ~.~~~~~~~~~~~~~(\varepsilon<<1)
\label{increase}
\eeq
with $\epsilon = 1$ for $\beta^{(0)}_{21} = 0+\varepsilon$ and $\epsilon =
-1$ for $\beta^{(0)}_{21} = \pi +\varepsilon$. Consider a vortex
configuration of type $(1,1)$ where the zeroes of each Higgs field are well
separated.  Then, the above surplus of energy is positive for $\epsilon = 
1$ and decreases as the cores are taken further appart and the overlap
diminishes hence the interaction in this case is repulsive. When perturbing
around the anti-aligned case, $\beta^{(0)}_{21} = \pi+\varepsilon$, the
energy increment (\ref{increase}) reverses sign. Previously non-interacting,
$(1,-1)$ antiparallel vortex configurations tend to increase the
overlap in order to lower the perturbation, and hence the force is
attractive. 

In summary, when perturbing around the aligned or misaligned scenarios, 
the vortex configurations are not  neutral anymore, and the interactions 
follow the pattern that was previously named ``hybrid type II'' where, if
made of distinct U(1)'s,  parallel vortices repel and antiparallel vortices
attract.

%%%%%%%%%%%%%%%%%%%%%%%%%%%%%%%%%%%%%%%%%%%%%%%%%%%%%%%%%%%%%%%%%%%%%%%%%%%%%%%
%%%%%%%%%%%                         Section VII                     %%%%%%%%%%%
%%%%%%%%%%%%%%%%%%%%%%%%%%%%%%%%%%%%%%%%%%%%%%%%%%%%%%%%%%%%%%%%%%%%%%%%%%%%%%%
\section{\bf Half Higgsed Vacua}

As pointed out in \cite{AD}, for particular values of the soft breaking
parameters $\mu$ and $\nu$ we have four instead of five  vacua. This happens
whenever one of the two half Higgsed vacua $\{a^D_{1} \neq 0, a^D_{2} =
0\}$ with  $C_1(\mu,\nu) = 0$,  or ($1\leftrightarrow 2$), meets and replaces
one of the normal vacua at $\{a^D_{1} = 0, a^D_{2} = 0\} $. This possibility
is actually achieved by turning off $C_i^{(0)}$ for $i=1$ or $2$. Since
precisely at the $Z_2$ point we have (\ref{cesu}), this amounts to $\mu$ and
$\nu$ fulfilling $\mu  = \mp \nu \Lambda$.
Let us choose for definiteness, $C_2^{(0)} = 0$. Inserting this back into
(\ref{pottbose}), the effective potential at the maximal point reads
\beq
V = {1\over 8} \lambda_1 (|\varphi_1|^2-v_1^2)^2 +  {1\over 8} \lambda_2
|\varphi_2|^4 - {1\over 4} \gamma \cos\beta_1 |\varphi_2|^2
(|\varphi_1|^2-v_1^2)~.
\label{super}
\eeq
Observe that the phase of $\varphi_2$ is free. When $\cos\beta_1<0$ this is
precisely the type of situation that was studied by Witten \cite{Witten}
and shown to lead to superconducting strings for specific ranges of
parameters. Let us briefly recall the essence of the mechanism. As the vacuum
equations (\ref{minim1}) exhibit, only the first $U(1)$  is
broken by the v.e.v. $\langle \varphi_1 \rangle = v_1$, whereas the second
$U(1)$  remains intact since $\langle\varphi_2\rangle = 0$.
This is fine for vacuum solutions, but suppose now that $\varphi_1$ developes
a vortex line. At the core of the vortex $\langle\varphi_1\rangle = 0$ and, in
turn, it may become favorable that $\langle\varphi_2\rangle \neq 0$ there.
Actually the model considered in \cite{Witten} is slightly more general than
ours involving the potential
\beq
V =   {1\over 8} g (|\varphi_1|^2-v^2)^2 + {1\over 4} \tilde g |\varphi_2|^4
+ f|\varphi_1|^2 |\varphi_2|^2 - m^2 |\varphi_2|^2 ~.
\label{potwitt}
\eeq
The detailed analysis of the dynamics showed that for parameters  in the
range $fv^2 \geq m^2$,  instability actually takes over and the
superconducting string indeed forms. We see easily that the present situation
lies precisely at the boundary of the region of validity, since in our case
$fv^2-m^2  = 0$, and the induced mass term for $\varphi_2$ exactly vanishes.
In \cite{springs}, this situation was also studied and seen to yield a power
law decay of the profile of $\varphi_2$ which leads to a long range scalar
attractive interaction among vortices.

At this point we would not like to put forward too strong a claim, but
simply point out the ocurrence of this coincidence among models. The possible
existence and relevance of structures like superconducting strings in the
microscopic context of confinement models should be handled with care.  For
example the question of quantum tunnelling will be certainly much more
relevant here than for cosmic strings.  Incidentally this question 
was also addressed in \cite{springs} where  it was seen that these power law
solutions are more stable than the usual ones.

As compared with  Witten's model, the one here involves the additional feature
of the non-diagonal kinetic term for the (dual) vector particles ({\em cf.} eq.
(\ref{efflagdos})). But precisely the fact that the quadratic forms of kinetic
term and potential are related paves the way to the  possibility of
rewritting the energy as a sum of squares (\ref{bogodos}). We may therefore 
expect vortex solutions of the superconducting type with dynamical properties
of BPS configurations. We can check that this is indeed the case by looking
at the smooth deformation of a generic (anti-)aligned   scenario.
\footnote{~As we approach the situation when $C_2 \to 0$, the parameters that
enter (\ref{super}) are such that $ \gamma,\lambda_2 << \lambda_1$ (see
eqns.(\ref{coup}) and (\ref{lyg1})--(\ref{lyg2})).  Hence at very low energy
the second $U(1)$ seemingly decouples. This is suggested by the $\N=2$ exact
effective solution, although it is reasonable to expect modifications of the
renormalization group flow in the
$\N=1$ theory.} Let us follow a continuous line of anti-aligned
($\beta_{21}^{(0)} =
\pi$) vacua 
$C_1^{(0)}\neq 0, C_2^{(0)} \to 0$. Precisely in this situation,
(\ref{ddeun}) presents no obstruction to a smooth deformation of the $(n,0)$
solutions down to the situation where $v_2 = 0$. In this limit the profiles of
$|\varphi_1|$ and $|\varphi_2|$  are correlated in such a way that   both
vanish at opposite ends. In fact, as $|\varphi_1|^2$ varies from zero up to
$v_1^2$ far away,  
$|\varphi_2|$ interpolates between
$\frac{\gamma}{\lambda_2}v_1^2= (\log 2/8\pi^2)
g_{D,1}^2 v_1^2$ at the origin (which need not be small !), and $0$ at
infinity.    
Moreover, since the phase of
$\varphi_2$ is free, the same arguments of ref.  \cite{Witten} can be used to
show that a persistent  current occurs. We  would call this a {\em BPS
superconducting string solution}.
%%%%%%%%%%%%%%%%%%%%%%%%%%%%%%%%%%%%%%%%%%%%%%%%%%%%%%%%%%%%%%%%%%%%%%%%%%%%%%%
%%%%%%%%%%%                         Section VIII                   %%%%%%%%%%%%
%%%%%%%%%%%%%%%%%%%%%%%%%%%%%%%%%%%%%%%%%%%%%%%%%%%%%%%%%%%%%%%%%%%%%%%%%%%%%%%
\section{Concluding Remarks}

The present paper is devoted to the low energy dynamics of ${\cal N}=2$
supersymmetric gauge theories softly broken to ${\cal N}=1$ by a
superpotential containing up to cubic perturbations. The effective lagrangian
in the neighborhood of maximal singularities of the quantum moduli space
corresponds to an Abelian $U(1)^{N-1}$ multi Higgs system with couplings
among different dual $U(1)$ factors. The case of $SU(3)$ has been analized in
some detail. There are generically no BPS electric vortices in the system
unless the soft breaking parameters have coincident complex phases (or they
differ by $\pi$) and the ultrastrong scaling limit \cite{HSZ} is taken. We
have seen that the effect over a BPS configuration of turning on an
infinitesimal misalignment among these parameters is the appearance of a net
repulsive force between parallel vortices corresponding to (zeroes of)
different Higgs fields. In a finite piece of material, metastable solutions
take place and vortices develope  static triangular lattice. We call this
phase ``hybrid Type II'' dual superconductivity.

When the theory is perturbed with a cubic superpotential, the ratio of string
tensions differs from that computed in the quadratic case \cite{DS} both when
the $\Tr\Phi^2$ perturbation is present or not. In the former case, we found
that these ratios even depend on the supersymmetry breaking parameters. These
results were obtained after imposing the ultrastrong scaling limit. It would
be certainly interesting to know if similar results emerge in the context of
MQCD. This is intriguing in the sense that string tensions in MQCD are given
by the distance of D4-branes which, for a single Casimir perturbation, are
stretched at the roots of unity over a circle of radius of order $\Lambda$
\cite{HSZ}, so one would not expect them to be modified (except, possibly, for
a global factor due to an induced change in $\Lambda$) as compared to the
purely quadratic case.

A natural extension of the present work involves   the case of
${\cal N}=2$ supersymmetric theories softly broken down to ${\cal N}=0$, and
possible soft breaking by higher than the two first Casimir operators.
This program can be addressed within the Whitham approach to the
Seiberg--Witten solution, where the   slow-times of the hierarchy 
can be used as spurionic sources of soft supersymmetry breaking \cite{emm}.
\bigskip

\acknowledgments
We are pleased to thank Jos\'e F. Barb\'on, A. Gonz\'alez-Arroyo, Michael
Douglas, Amihay Hanany and Marcos Mari\~no  for interesting discussions.
J.M. wants to thank J.J. Blanco Pillado for pointing out reference
\cite{springs}. The work of J.D.E. has been supported by the National
Research Council (CONICET) of Argentina and the Ministry of Education and
Culture of Spain. The work of J.M. was partially supported by DGCIYT under
contract PB96-0960 and European Union TMR grant ERBFM-RXCT960012.

%---------------- Bibliografia-------------------

\end{document}